\documentstyle[twoside,psfig]{article}
\catcode`\@=11
\long\def\@makefntext#1{
\protect\noindent \hbox to 3.2pt {\hskip-.9pt
$^{{\eightrm\@thefnmark}}$\hfil}#1\hfill}		

\def\@makefnmark{\hbox to 0pt{$^{\@thefnmark}$\hss}}	

\def\ps@myheadings{\let\@mkboth\@gobbletwo
\def\@oddhead{\hbox{}
\rightmark\hfil\eightrm\thepage}
\def\@oddfoot{}\def\@evenhead{\eightrm\thepage\hfil
\leftmark\hbox{}}\def\@evenfoot{}
\def\sectionmark##1{}\def\subsectionmark##1{}}

\oddsidemargin=\evensidemargin
\newcounter{sectionc}\newcounter{subsectionc}\newcounter{subsubsectionc}
\renewcommand{\section}[1] {\vspace{12pt}\addtocounter{sectionc}{1}
\setcounter{subsectionc}{0}\setcounter{subsubsectionc}{0}\noindent
	{\tenbf\thesectionc. #1}\par\vspace{5pt}}
\renewcommand{\subsection}[1] {\vspace{12pt}\addtocounter{subsectionc}{1}
	\setcounter{subsubsectionc}{0}\noindent
	{\bf\thesectionc.\thesubsectionc. {\kern1pt \bfit #1}}\par\vspace{5pt}}
\renewcommand{\subsubsection}[1] {\vspace{12pt}\addtocounter{subsubsectionc}{1}
	\noindent{\tenrm\thesectionc.\thesubsectionc.\thesubsubsectionc.
	{\kern1pt \tenit #1}}\par\vspace{5pt}}
\newcommand{\nonumsection}[1] {\vspace{12pt}\noindent{\tenbf #1}
	\par\vspace{5pt}}
\newcounter{appendixc}
\newcounter{subappendixc}[appendixc]
\newcounter{subsubappendixc}[subappendixc]
\renewcommand{\thesubappendixc}{\Alph{appendixc}.\arabic{subappendixc}}
\renewcommand{\thesubsubappendixc}
	{\Alph{appendixc}.\arabic{subappendixc}.\arabic{subsubappendixc}}

\renewcommand{\appendix}[1] {\vspace{12pt}
        \refstepcounter{appendixc}
        \setcounter{figure}{0}
        \setcounter{table}{0}
        \setcounter{lemma}{0}
        \setcounter{theorem}{0}
        \setcounter{corollary}{0}
        \setcounter{definition}{0}
        \setcounter{equation}{0}
        \renewcommand{\thefigure}{\Alph{appendixc}.\arabic{figure}}
        \renewcommand{\thetable}{\Alph{appendixc}.\arabic{table}}
        \renewcommand{\theappendixc}{\Alph{appendixc}}
        \renewcommand{\thelemma}{\Alph{appendixc}.\arabic{lemma}}
        \renewcommand{\thetheorem}{\Alph{appendixc}.\arabic{theorem}}
        \renewcommand{\thedefinition}{\Alph{appendixc}.\arabic{definition}}
        \renewcommand{\thecorollary}{\Alph{appendixc}.\arabic{corollary}}
        \noindent{\tenbf Appendix \theappendixc #1}\par\vspace{5pt}}
\newcommand{\subappendix}[1] {\vspace{12pt}
        \refstepcounter{subappendixc}
        \noindent{\bf Appendix \thesubappendixc. {\kern1pt \bfit #1}}
	\par\vspace{5pt}}
\newcommand{\subsubappendix}[1] {\vspace{12pt}
        \refstepcounter{subsubappendixc}
        \noindent{\rm Appendix \thesubsubappendixc. {\kern1pt \tenit #1}}
	\par\vspace{5pt}}

\newcommand{\textlineskip}{\baselineskip=13pt}
\newcommand{\smalllineskip}{\baselineskip=10pt}

\def\eightcirc{
\begin{picture}(0,0)
\put(4.4,1.8){\circle{6.5}}
\end{picture}}
\def\eightcopyright{\eightcirc\kern2.7pt\hbox{\eightrm c}}

\newcommand{\copyrightheading}[1]
	{\vspace*{-2.5cm}\smalllineskip{\flushleft
	{\footnotesize International Journal of Modern Physics C #1}\\
	{\footnotesize $\eightcopyright$\, World Scientific Publishing
	 Company}\\
	 }}

\newcommand{\publisher}[2]{{\begin{center}\footnotesize\smalllineskip
	Received #1\\
	Revised #2
	\end{center}
	}}

\def\abstracts#1#2#3{{
	\centering{\begin{minipage}{4.5in}\footnotesize\baselineskip=10pt
	\parindent=0pt #1\par
	\parindent=15pt #2\par
	\parindent=15pt #3
	\end{minipage}}\par}}

\def\keywords#1{{
	\centering{\begin{minipage}{4.5in}\footnotesize\baselineskip=10pt
	{\footnotesize\it Keywords}\/: #1
	\end{minipage}}\par}}

\newcommand{\bibit}{\nineit}
\newcommand{\bibbf}{\ninebf}
\renewenvironment{thebibliography}[1]
        {\frenchspacing
	 \ninerm\baselineskip=11pt
         \begin{list}{\arabic{enumi}.}
        {\usecounter{enumi}\setlength{\parsep}{0pt}
	 \setlength{\leftmargin 12.7pt}{\rightmargin 0pt} 
         \setlength{\itemsep}{0pt} \settowidth
	{\labelwidth}{#1.}\sloppy}}{\end{list}}

\newcounter{itemlistc}
\newcounter{romanlistc}
\newcounter{alphlistc}
\newcounter{arabiclistc}

\newcommand{\fcaption}[1]{
        \refstepcounter{figure}
        \setbox\@tempboxa = \hbox{\footnotesize Fig.~\thefigure. #1}
        \ifdim \wd\@tempboxa > 5in
           {\begin{center}
        \parbox{5in}{\footnotesize\smalllineskip Fig.~\thefigure. #1}
            \end{center}}
        \else
             {\begin{center}
             {\footnotesize Fig.~\thefigure. #1}
              \end{center}}
        \fi}

\newcommand{\tcaption}[1]{
        \refstepcounter{table}
        \setbox\@tempboxa = \hbox{\footnotesize Table~\thetable. #1}
        \ifdim \wd\@tempboxa > 5in
           {\begin{center}
        \parbox{5in}{\footnotesize\smalllineskip Table~\thetable. #1}
            \end{center}}
        \else
             {\begin{center}
             {\footnotesize Table~\thetable. #1}
              \end{center}}
        \fi}

\def\@citex[#1]#2{\if@filesw\immediate\write\@auxout
	{\string\citation{#2}}\fi
\def\@citea{}\@cite{\@for\@citeb:=#2\do
	{\@citea\def\@citea{,}\@ifundefined
	{b@\@citeb}{{\bf ?}\@warning
	{Citation `\@citeb' on page \thepage \space undefined}}
	{\csname b@\@citeb\endcsname}}}{#1}}

\newif\if@cghi
\def\cite{\@cghitrue\@ifnextchar [{\@tempswatrue
	\@citex}{\@tempswafalse\@citex[]}}
\def\citelow{\@cghifalse\@ifnextchar [{\@tempswatrue
	\@citex}{\@tempswafalse\@citex[]}}
\def\@cite#1#2{{$\null^{#1}$\if@tempswa\typeout
	{IJCGA warning: optional citation argument
	ignored: `#2'} \fi}}

\def\pmb#1{\setbox0=\hbox{#1}
	\kern-.025em\copy0\kern-\wd0
	\kern.05em\copy0\kern-\wd0
	\kern-.025em\raise.0433em\box0}

\def\fnt#1#2{\footnotetext{\kern-.3em
	{$^{\mbox{\scriptsize #1}}$}{#2}}}

\def\ps@myheadings{%
    \let\@oddfoot\@empty\let\@evenfoot\@empty
    \def\@evenhead{\slshape\leftmark\hfil}
    \def\@oddhead{\hfil{\slshape\rightmark}}
    \let\@mkboth\@gobbletwo
    \let\sectionmark\@gobble
    \let\subsectionmark\@gobble
    }
\font\tenrm=cmr10
\font\tenit=cmti10
\font\tenbf=cmbx10
\font\bfit=cmbxti10 at 10pt
\font\ninerm=cmr9
\font\nineit=cmti9
\font\ninebf=cmbx9
\font\eightrm=cmr8

\def\qed{\hbox{${\vcenter{\vbox{		    
   \hrule height 0.4pt\hbox{\vrule width 0.4pt height 6pt
   \kern5pt\vrule width 0.4pt}\hrule height 0.4pt}}}$}}

\def\bsc{{\sc a\kern-6.4pt\sc a\kern-6.4pt\sc a}}  	
\def\bflatex{\bf L\kern-.30em\raise.3ex\hbox{\bsc}\kern-.14em
T\kern-.1667em\lower.7ex\hbox{E}\kern-.125em X}

\begin{document}
\setlength{\textheight}{7.7truein}  

\thispagestyle{empty}

\markboth{\protect{\footnotesize\it Instructions for Typesetting
Manuscripts}}{\protect{\footnotesize\it Instructions for
Typesetting Manuscripts}}

\normalsize\textlineskip

\setcounter{page}{1}

\copyrightheading{}			

\vspace*{0.88truein}

\centerline{\bf Regulation effects on market with}
\vspace*{0.035truein}
\centerline{\bf Bak-Sneppen model in high dimensions}
\vspace*{0.37truein}
\centerline{\footnotesize Takuya Yamano\footnote{Department of
Applied Physics, Faculty of Science, Tokyo Institute of Technology,
Oh-okayama, Meguro-ku Tokyo, 152-8551, Japan\\
\it E-mail: tyamano@mikan.ap.titech.ac.jp}}
\baselineskip=12pt
\centerline{\footnotesize\it Institute for Theoretical Physics,
Cologne University, D-50923 K\"{o}ln, Euroland}
\centerline{\footnotesize\it E-mail: ty@thp.Uni-Koeln.DE}

\vspace*{0.225truein}
\publisher{(received date)}{(revised date)}

\vspace*{0.25truein}
\abstracts{We present the effect of regulations on self-organized
market by using biological model of Bak-Sneppen in higher dimensions. This
study extends the idea of Cuniberti et.al. The higher-dimensional
description of the market suffices less effect of regulation than that of
lower one.}{}{}

\vspace*{5pt}


\vspace*{1pt}\textlineskip	
\section{Introduction}		
\vspace*{-0.5pt}
\noindent

Economic markets are known to exhibit a self-organized criticality(SOC)
\cite{1}. The SOC is characterized by a spontaneous approach to a
steady state without any external tuning of parameters. However, there
seems to exist many external factors which affect on the constituent
companies in real economical markets such as a certain kind of control
of prices, a direct regulation to the companies and so forth.
Recently Cuniberti et. al.\cite{2} introduced a regulations on
the self-organized market with Bak-Sneppen model of one dimension.
The Bak-Sneppen model\cite{3,4} is
known as a simple model which exhibits SOC. Although the model
is originally devised to describe a evolutionary aspects of biological
species which have varied niches, it seems possible to borrow the concept
in order to describe the asymptotic market behavior by regarding species as
companies in the market. We follow Cuniberti et al \cite{2} and
are concerned with the effects of a regulation
from authorities such as governments or agencies on economic
markets. In this paper, we extend the market to $d$-dimensional lattices and
see the regulation effects for each dimension. The purpose of the present
simulation is not to justify the validity of the Bak-Sneppen model as a
self-organized market. Instead we aim at elucidating the nature of
regulations and seeing dependency on the dimensionality.

\section{Model}
\vspace*{-0.5pt}
\noindent
We assume that companies are located on sites in $d$-dimensional lattices.
Each company on the lattice is characterized by one single parameter
which is called fitness value $f_i$. The fitnesses are assumed to express a
certain kind of strength of the company\cite{2}.
For example, it may be the size of
the company, the amount of capital, the profit and so on. We consider that
a company whose fitness value is high can survive easily in a competitive
and highly interactive market. The closeness of sites on the lattice
represents that the companies are simular. The dynamics we
employ is as follows.
First we assign random numbers between $0$ and $1$ to each company as an
initial condition. The market is updated by finding the least-fit company
(the company having the lowest fitness value in the lattice) and assigning
new random numbers $f^\prime$ to it and to the $2d$ nearest neighbor
companies independently.
At each time step(e.g.fiscal year), a company with the lowest fitness value
may be forced to improve its performance or may go bankrupt and be
replaced by a new company. The reason
for incorporating the nearest neighbors is that the updated least-fit one
may make a new environment around it and the nearest neighbor companies
feels the change. Accordingly, a company with a high fitness does not update
on its own but might eventually be affected by an updating of the
neighbor company, causing either an improvement or a deterioration of
its fitness. As a result of updating their fitness
values, however, some may happen to get a very high fitnesses
(too rapid a growth causes problems\cite{2}).
Our implementation of a regulation to this dynamics is a following way.
At each time step, we identify $1+2d$ sites as updating sites.
After the first updating for all $1+2d$ sites, if some of the $1+2d$ site 
have $f^\prime > \eta$ ($\eta\in [0,1]$), all these sites are updated 
once again (new random numbers are assigned). However,
even if then the new fitness values for some of the $1+2d$ sites exceed $\eta$ 
again, we do not update them and the dynamics will proceed to the next time 
step. The regulation $\eta$ may be considered to be a boundary imposing 
high tax rates on the companies above $\eta$. This procedure completes 
one time step. We iterate this until the market reaches its critical state.

\section{Results}
\vspace*{-0.5pt}
\noindent
We performed the simulation up to 7 dimensions. We selected the lattice
size $L$ for each dimension $d$ as follows: $(d,L)$ =(1,20000),
(2,200), (3,40), (4,13), (5,9), (6,6), (7,4). To achieve steady state, we
iterated each step $10^8$ times for one dimension and $10^7$ times for
the other dimensions. The consistent results are obtained with the
previous paper\cite{2} for all dimensions in the histogram of the fitness
(probability density distribution). As an example, we show the histogram
of the fitness for the three- dimensional case in Fig.1. The application of the
regulation lowers the location of the discontinuity $f_c$ and the number of
companies having high fitnesses larger than $\eta$ is reduced.
We see, on the other hand, that the companies having lower fitness
can stay in market thanks to the regulation\cite{2}.

\noindent
\begin{figure}[htbp]
\vspace*{13pt}
\centerline{\psfig{file=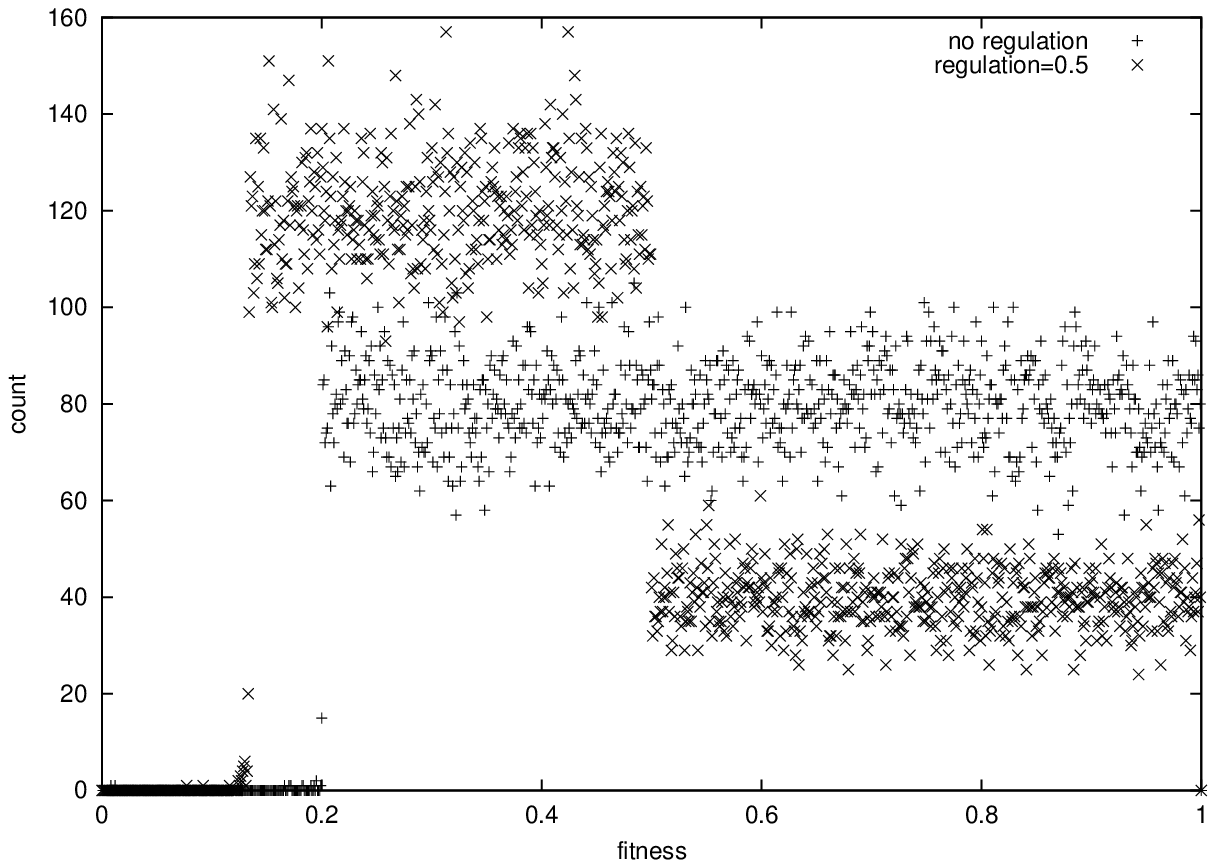}} 
\vspace*{13pt}
\fcaption{Histogram of the fitness in 3 dimension with $L=40$.
Crosses and pluses shows the no regulation on the market and a regulation
,$\eta=0.5$, respectively.}
\end{figure}

In a laissez-faire market without no regulations
($\eta =0$), we plotted
the $f_c$ against the dimension in the log-log scale in Fig.2.
The straight line on the plot shows a power law property, $f_c\sim d^{-1.1}$.

\noindent
\begin{figure}[htbp]
\vspace*{13pt}
\centerline{\psfig{file=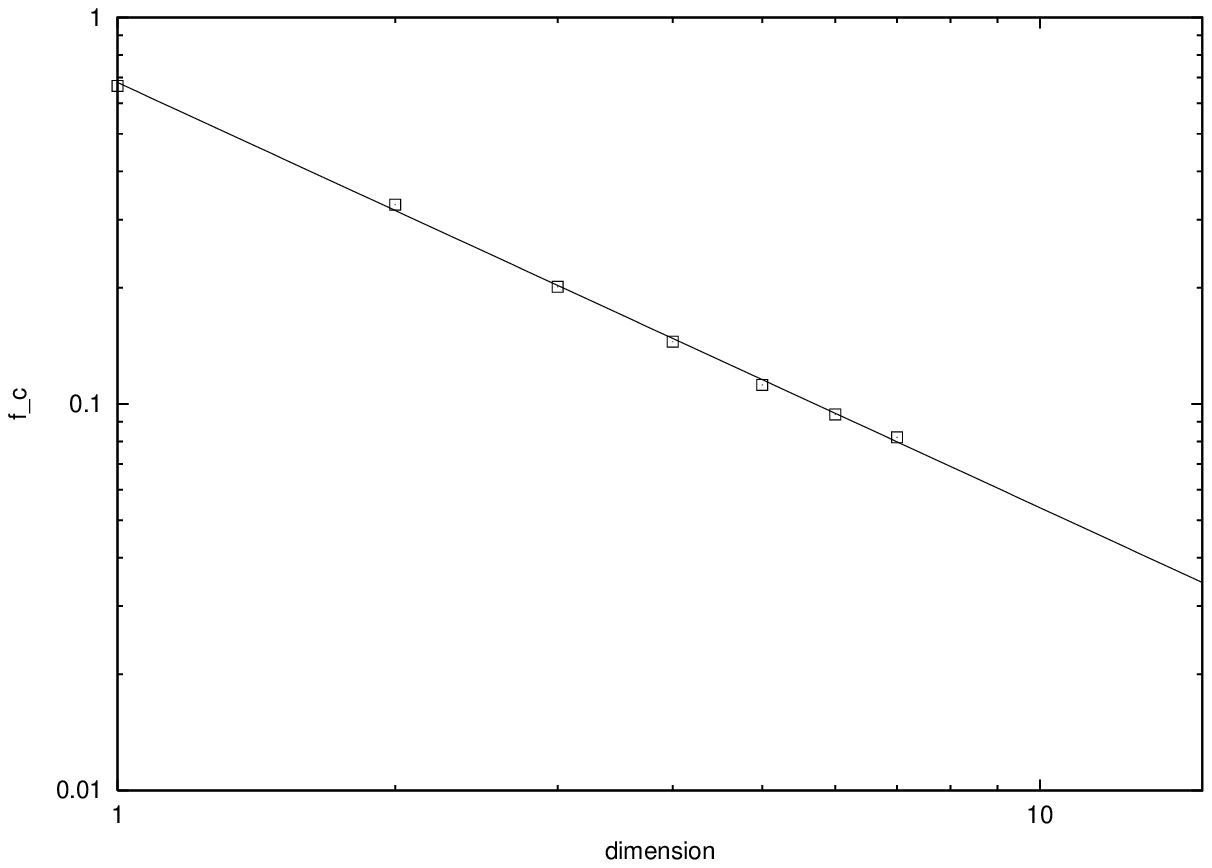}} 
\vspace*{13pt}
\fcaption{The critical fitness against the dimension.
Since the straight line on log-log plot means a power low, we find the
relation $f_c\sim d^{-1.1}$.}
\end{figure}

Fig.3 shows the effect of regulations on critical fitness value.
We can see the market with higher dimension shows less effect of the
regulation than that of lower dimension. Moreover there is a most efficient
value of $\eta$ lowering the critical fitness value $\eta_{eff}$.
The $\eta_{eff}$ shifts towards small values as the market dimension
increases.

\noindent
\begin{figure}[htbp]
\vspace*{13pt}
\centerline{\psfig{file=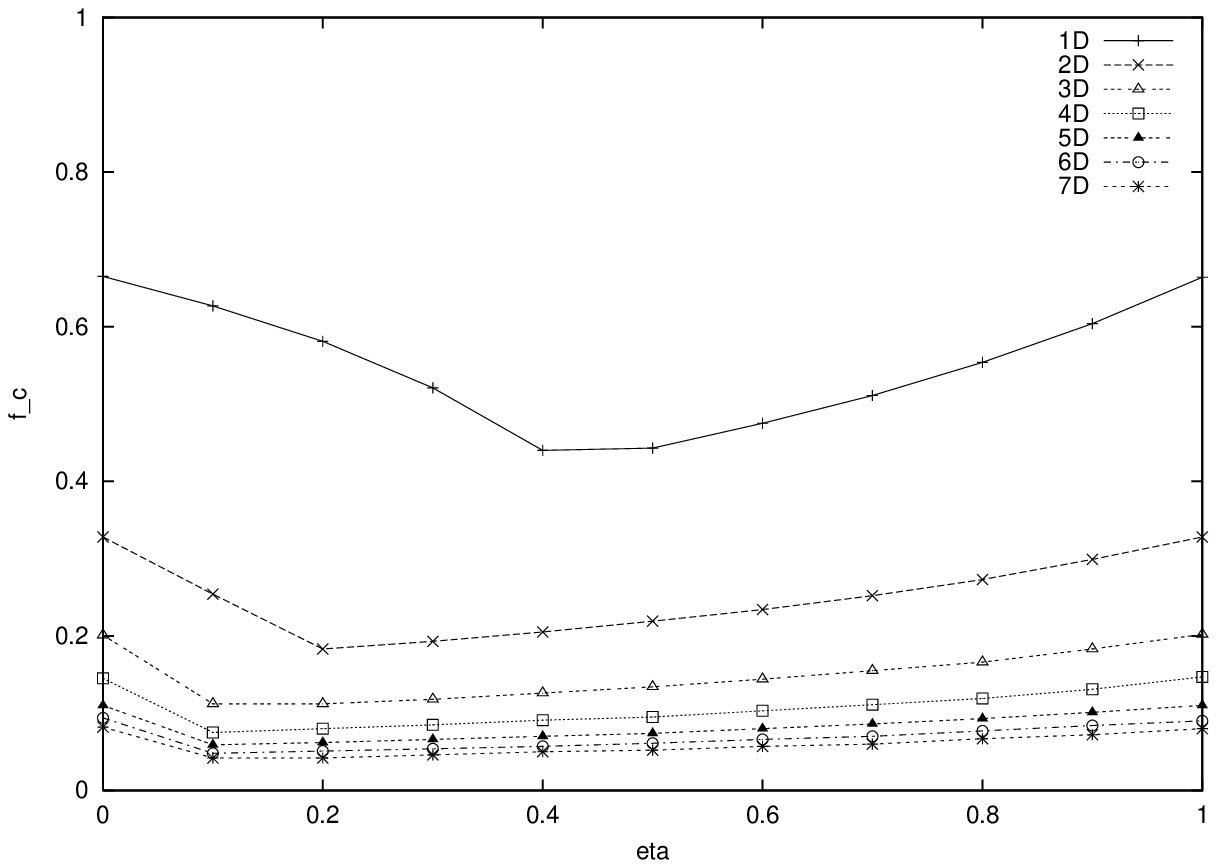}} 
\vspace*{13pt}
\fcaption{The critical fitness value against the regulation
for $d$-dimensional market. The regulation lowers the $f_c$ in all dimensions.}
\end{figure}

\section{Summary}
\vspace*{-0.5pt}
\noindent
We have simulated the Bak-Sneppen model in higher dimensions to see
the effect of control on the self-organized market. Previously,
Ray and Jan\cite{5} simulated the Bak-Sneppen model up to 4 dimensions
in a different context. However the relation between the critical fitness
and dimensions was not given there. We found a simple power-law relation
between them $f_c\sim d^{-1.1}$. Furthermore, in the description of the market
in higher dimensions, the market allows companies to have a wider range of
their fitnesses and we may say that the application of regulations
have not so much effects on the market.  \\

\nonumsection{Acknowledgements}
\noindent
We acknowledge the financial support from
 the DAAD (Deutscher Akademischer Austauschdienst) for staying the
Universit\"{a}t zu K\"{o}ln, where this work was suggested by D. Stauffer.

\end{document}